\documentclass[12pt]{article}

\usepackage{amsfonts}
\usepackage{amssymb}
\usepackage{amscd}

\def\D{\hbox{D\kern-.73em\raise.25ex\hbox{-}\raise-.25ex\hbox{ }}}
 \def\d{\hbox{d\kern-.33em\raise.75ex\hbox{-}\raise-.75ex\hbox{}}}

\def\GGG{\frak G }
\def\gr3{\GGG\,(\SSS_3)}

\def\gr2{\GGG\,(\SSS_2)}

\def\SSS{\frak S}

\def\ed{\end{document}}
\def\beq{\begin{equation}}
\def\eeq{\end{equation}}
\def\bea{\begin{eqnarray}}
\def\eea{\end{eqnarray}}
\def\ba{\begin{array}}
\def\ea{\end{array}}
\def\bi{\begin{itemize}}
\def\ei{\end{itemize}}

\def\noi{\noindent}

\newcommand{\bp}{\noindent\begin{minipage}[c]}
\newcommand{\ep}{\end{minipage}}

\begin{document}
 \baselineskip=11pt

\title{\bf \Large Linear Fractional $p$-Adic and Adelic \\
  Dynamical Systems\hspace{.25mm} }
\author{{Branko Dragovich}\hspace{.25mm}\thanks{\,e-mail
address: dragovich@phy.bg.ac.yu}
\\ \normalsize{Institute of Physics, P.O. Box 57, 11001 Belgrade, Serbia}
\vspace{2mm}
\\ {Andrei Khrennikov} \hspace{.25mm} \\ \normalsize{International Center for Mathematical
Modeling} \\ \normalsize{ MSI, V\"axj\"o University,
SE-35195, V\"axj\"o, Sweden }  \vspace{2mm} \\
{Du\v san Mihajlovi\'c}\hspace{.25mm}
\\ \normalsize{Faculty of Physics, P.O. Box 368, 11001 Belgrade, Serbia}}

\date{}

\maketitle


\begin{abstract}
Using an adelic approach we simultaneously consider real and
$p$-adic aspects of dynamical systems whose states are mapped by
linear fractional transformations isomorphic to some subgroups of
$GL (2, \mathbb{Q})$, $SL (2, \mathbb{Q})$ and $SL (2, \mathbb{Z})$
groups. In particular, we investigate behavior of these adelic
systems when fixed points are rational. It is shown that any of
these rational fixed points is $p$-adic indifferent for all but a
finite set of primes. Thus only for finite number of $p$-adic cases
a rational fixed point may be attractive or repelling. It is also
shown that real and $p$-adic norms of any nonzero rational fixed
point are connected by adelic product formula.

\end{abstract}

\bigskip

PACS numbers: 05.45.-a , 02.10.De

Keywords:  $p$-adic dynamics, adelic dynamics, fixed points
\bigskip

\bigskip

\bigskip

\section{\large Introduction }

There are many dynamical systems whose states change in discrete
time intervals. In such discrete time dynamical system its state
changes by  a mapping
 \beq         f : X \rightarrow X ,  \label{1.1} \eeq
where $X$ is the state space and the map $f$ describes how states
evolve in time units. It is suitable to study evolution of such time
discrete systems by iteration.  If the state at the time $t=0$ is
$x_{0} \in X$ and $f^{n}= f \circ \cdots \circ f$ then after $n$
iterations the state becomes \beq x_{n} = f^{n}(x_{0}) .
\label{1.2}\eeq

The state space $X$ has usually some natural additional structures
such as hierarchies and distances between states. In particular, in
physics and other related topics, the state space of very complex
systems often displays a hierarchical structure. This implies that
the classification of the states and their relationships may be
based on an ultrametric, and in particular $p$-adic distance $d_p$.
Recently much attention has been paid to some $p$-adic dynamical
systems, since they have a lot of potential applications (for a
review, see \cite{khrennikov4}).

Among the disordered systems, the mean field models for spin glasses
whose ground states have ultrametric structure are of particular
interest in mathematical physics \cite{parisi1}. Methods of p-adic
analysis are applied to the investigation of replica symmetry
breaking \cite{parisi2}. The ultrametricity  arises from the basic
properties of the field of p-adic numbers, the most important
example of ultrametric spaces. This p-adic reformulation and further
generalizations could be useful starting point in the study of the
whole structure of spin glasses \cite{khrennikov2}.

Also the cut and project method, commonly used in the study of
quasicrystals and aperiodic order, has been recently extended
\cite{moody}. Namely the key ingredients od this scheme, internal
spaces, are no longer Euclidean but  spaces with non-Euclidean
topologies. Namely $p$-adic topologies or mixed Euclidean/$p$-adic
topologies are combined in the physical-internal space pair.

From the above examples it seems that ultrametricity is a common
ingredient which cannot be avoided, and its $p$-adic treatment seems
to be quite natural. Since 1987 there have been many constructions
of $p$-adic physical models. In particular, $p$-adic numbers have
been successfully  employed in string theory, quantum mechanics and
quantum cosmology (for a review, see \cite{freund},
\cite{dragovich1} and \cite{dragovich2}, respectively).

When each state of physical system is associated to several distinct
hierarchical structures, the labeling of states by $p$-adics is no
longer sufficient and appropriate index set becomes the ring of
adeles. This was demonstrated for an asymmetric stochastic process
on the adeles \cite{karwowski}. Ingredients of adeles are real and
all $p$-adic numbers, which contain rationals in a dense way.
According to \cite{roth} (see also motivations in \cite{karwowski})
rational numbers imply foundation of quantum theory on adelic spaces
and some progress has been achieved in string theory, quantum
mechanics and quantum cosmology (for a review, see \cite{freund},
\cite{dragovich1} and \cite{dragovich2}, respectively).

Especially the unified aspect of adeles is  a motivation to employ
adelic approach to dynamical systems. In particular, we reconsider a
class of the rational $p$-adic dynamical systems \cite{mukhamedov1},
\cite{mukhamedov2} and construct the corresponding adelic dynamics
with rational fixed points. We expect that this adelic approach will
imply further developments of both $p$-adic and real dynamical
systems.

In our case the state space $X$ of the system is the adelic one and
discrete dynamics is described by the mapping \beq f (x) = \frac{a x
+ b}{c x + d}\,,  \label{1.3}\eeq where $a, b, c, d$ are some
rational numbers satisfying $a d - b c \neq 0$ and especially $a d -
b c =  1$ . It is worth noting that taking  physical parameters to
be rational numbers is not unnatural restriction but moreover gives
a possibility to treat real and $p$-adic properties simultaneously
and on an equal footing. In this way our approach differs from that
in \cite{mukhamedov1}, where parameters  $a\,, b\,, c\,, d \in
\mathbb{C}_p$.

Linear fractional transformations (M\"obius transformations)
(\ref{1.3}) and related $G L (2, \mathbb{C})$, $G L (2,
\mathbb{C}_p)$ groups, and their subgroups, have very rich
mathematical structure. They also have important applications in
many parts of theoretical physics (see, e.g. \cite{freund} and
\cite{dolan}).

Let us also mention that if in this dynamical system we let
parameters a,b,c,d to be sequences then we obtain recursion equation
for magnetization arising in the study of the spin systems on the
Bethe lattice. This implies  that adelic linear fractional dynamical
systems could be extended to the dynamics of spin systems on the
random tree-like graphs over $p$-adic numbers, random rational
$p$-adic systems and rationally perturbed monomial systems
\cite{mukhamedov3}.

The paper is organized as follows. In Sec. 2 we briefly present some
pertinent properties of $p$-adic numbers and adeles.  Sec. 3 is
devoted to an analysis in detail of real, $p$-adic and adelic
properties of the above linear fractional dynamics (\ref{1.3}).
Obtained results are slightly extended, discussed and summarized in
Sec. 4.

\section{\large $p$-Adic Numbers and Adeles}

Rational numbers are significant in physics as well as in
mathematics. Physical significance comes from the fact that result
of any measurement is a rational number. According to the Ostrowski
theorem, the set $\mathbb{Q}$ of rational numbers is a dense
subfield not only in the field $\mathbb{R}$ of real numbers but also
in the field $\mathbb{Q}_p$ of $p$-adic numbers, for every prime
number $p$. Consequently, a research on the basis of $\mathbb{R}$
and $\mathbb{Q}_p$ provides rather fine description of many
properties related to rational numbers. The space of adeles
$\mathbb{A}$ is a suitable tool to consider $\mathbb{R}$ and
$\mathbb{Q}_p$ simultaneously and on the equal footing. Below we
shall give a short introductory review of $p$-adic numbers and
adeles relevant for this paper.

Let us recall  that the first infinite set of numbers we encounter
is the set ${\mathbb N}$ of natural numbers. To have a solution of
the simple linear equation $ x + a = b$ for any $a, b \in {\mathbb
N}$, one has to extend ${\mathbb N}$ and to introduce the set
${\mathbb Z}$ of integers. Requiring that there exists solution of
the linear equation $ n x = m$ for any $0\neq n, m \in {\mathbb Z}$
one obtains the set ${\mathbb Q}$ of rational numbers. Evidently
these sets satisfy ${\mathbb N}\subset {\mathbb Z}\subset {\mathbb
Q}$. Algebraically ${\mathbb N}$ is a semigroup, ${\mathbb Z}$ is a
ring, and ${\mathbb Q}$ is a field.

To get ${\mathbb Q}$ from ${\mathbb N}$ only algebraic operations
are used, but to obtain the field ${\mathbb R}$ of real numbers from
${\mathbb Q}$ one has to employ the absolute value which is an
example of the norm (valuation) on ${\mathbb Q}$. Let us recall that
a norm on ${\mathbb Q}$ is a map $||\cdot ||: {\mathbb Q}\to
{\mathbb R_{+}} = \{ x\in {\mathbb R}\,:\,\,  x\geq 0 \} $ with the
following properties: (i) $||x|| =0 \leftrightarrow x=0$, (ii)
$||x\cdot y|| = ||x||\cdot ||y|| $, and $\, ||x +y ||\, \leq \,
||x|| + ||y|| \,$ for all $x, y \in {\mathbb Q}$. In addition to the
absolute value, for which  we  use usual arithmetic notation
$|\cdot|_\infty$, one can introduce on ${\mathbb Q}$ a norm with
respect to each prime number $p$. Note that, due to the
factorization of integers, any rational number can be uniquely
written as $x = p^\nu \, \frac{m}{n}$, where $p,\, m,\, n$ are
mutually prime and $\nu \in {\mathbb Z}$. Then by definition
$p$-adic norm (or, in other words, $p$-adic absolute value) is $|
x|_p = p^{-\nu}$ if $ x \neq 0$ and $|0|_p =0$. One can verify that
$|\cdot |_p$ satisfies all the above conditions and, moreover,
strong triangle inequality, i.e. $|x + y|_p \leq \,  \mbox{max} \,
(|x|_p,\, |y|_p) $. Thus $p$-adic norms belong to the class of
non-Archimedean (ultrametric) norms. Up to the equivalence,  there
is only one  $p$-adic norm for every prime number $p$. According to
the Ostrowski theorem any nontrivial norm on ${\mathbb Q}$ is
equivalent either to the $|\cdot|_\infty$ or to one of the
$|\cdot|_p$. One can easily show that $|m|_p \leq 1$ for any $m \in
{\mathbb Z}$ and any prime $p$. The $p$-adic norm is a measure of
divisibility of the integer $m$ by prime $p$: the more divisible,
the $p$-adic smaller.  Using Cauchy sequences of rational numbers
one can make completions of ${\mathbb Q}$ to obtain ${\mathbb R}
\equiv {\mathbb Q}_\infty$ and the fields ${\mathbb Q}_p$ of
$p$-adic numbers using norms $|\cdot|_\infty$ and $|\cdot|_p \,$,
respectively. The cardinality of ${\mathbb Q}_p$ is the continuum,
like of ${\mathbb Q}_\infty$. $p$-Adic completion of ${\mathbb N}$
gives the ring ${\mathbb Z}_p = \{ x \in {\mathbb Q}_p :\,\, |x|_p
\leq 1 \}$ of $p$-adic integers. Denote by ${\mathbb U}_p = \{ x \in
{\mathbb Q}_p :\,\, |x|_p = 1 \}$ multiplicative group of $p$-adic
units.

Any $p$-adic number $ x \in {\mathbb Q}_p$ can be presented in the
unique way (unlike real numbers) as the sum of $p$-adic convergent
series of the form
\begin{equation}
x = p^\nu \, (x_0 + x_1 p  + \cdots + x_n p^n + \cdots ) , \quad \nu
\in {\mathbb Z} , \quad x_n \in \{0, 1, \cdots, p-1 \} . \label{2.1}
\end{equation}
It resembles representation of a real number $y= \pm\, 10^\mu \,
\sum_{k=0}^{-\infty} b_k 10^k ,\, \, \mu \in {\mathbb Z} , \, \, b_k
\in \{0, 1, \cdots, 9  \}$ , but with the expansion in the opposite
way. If $\nu \geq 0$ in (\ref{2.1}), then $x \in {\mathbb Z}_p$\,.
When $\nu = 0$ and $x_0 \neq 0$ one has $x \in {\mathbb U}_p$\,. Any
negative integer can be easily presented starting from the
representation of $- 1$:
\begin{equation}
- 1 = p-1 \ +\, (p-1) p\,  + \, (p-1) p^2 + \cdots + (p-1) p^n +
\cdots  . \label{2.2}
\end{equation}
Validity of (\ref{2.2}) can be shown by elementary arithmetics,
which is very similar to the real case, or treating it as the
$p$-adic convergent geometric series.

Using the norm $|\cdot|_p$ one can introduce $p$-adic metric $d_p
(x, y) = |x - y|_p $, which satisfies all necessary properties of
metric with strong triangle inequality, i.e. $d_p (x, y) \leq
\mbox{max} \,(\, d_p (x,z), \, d_p (z, y)\,)$ which is of the
non-Archimedean (ultrametric) form. Consequently $d_p (x,y)$ is a
distance between $p$-adic numbers $x$ and $y$. Using this metric,
${\mathbb Q}_p$ becomes an ultrametric space with $p$-adic topology.
 Because of ultrametricity, the $p$-adic spaces have some exotic (from the
real point of view) properties and usual illustrative examples are:
a) any point of the ball $B_\mu (a) = \{x \in {\mathbb Q}_p\, : \,
\,\, |x - a|_p \leq p^\mu \}$ can be taken as its center instead of
$a$; b) any ball can be regarded as a closed as well as an open set;
c) two balls may not have partial intersection, i.e. they are
disjoint sets or one of them is a subset of the other; and c) all
triangles are isosceles. ${\mathbb Q}_p$ is a zerodimensional and a
totally disconnected topological space. ${\mathbb Z}_p$ is a compact
and  ${\mathbb Q}_p$ is a locally compact space.

In analogy with real case one can introduce $p$-adic algebraic
extensions. Here the situation is much richer. The quadratic
equation \beq z^2 - \tau = 0\,, \quad \tau = p\,, \varepsilon\,,
\varepsilon p\,, \quad \mbox{where} \,\, \varepsilon = \sqrt[p-1]{1}
\, \mbox{and} \, p \neq 2 \,, \label{2.3} \eeq has not a solution in
$\mathbb{Q}_p$ and one must introduce quadratic extension
$\mathbb{Q}_p (\sqrt{\tau})$ with elements $z = x + \sqrt{\tau} \, y
\,, \quad x\,, y \in \mathbb{Q}_p \,. $ When $p =2$ there are seven
quadratic extensions $\mathbb{Q}_2 (\sqrt{\tau})$, where $\tau =
-1\,, \pm 2\,, \pm 3\,, \pm 6\,$. However $\mathbb{Q}_p
(\sqrt{\tau})$ are not algebraically closed and one has to  make
higher extensions. Algebraically closed and topologically complete
extension is $\mathbb{C}_p \,,$ which is an infinite dimensional
vector space. For a more details about $p$-adic numbers and their
algebraic extensions, see, e.g. \cite{schikhof}.

Real and $p$-adic numbers are continual extrapolations of rational
numbers along all possible nontrivial and inequivalent metrics. To
consider real and $p$-adic numbers simultaneously and on equal
footing one uses concept of adeles. An adele $x$ (see, e.g.
\cite{gelfand}) is an infinite sequence \beq
  x= (x_\infty\,, x_2\,, x_3\,, \cdots, x_p\,, \cdots), \label{2.4}\eeq
where $x_\infty \in {\mathbb R}$ and $x_p \in {\mathbb Q}_p$ with
the restriction that for all but a finite set $\mathcal P$ of primes
$p$ one has  $x_p \in {\mathbb Z}_p $. Componentwise addition and
multiplication make the ring structure of the set ${\mathbb A}$ of
all adeles, which is the union of restricted direct products in the
following form:
\begin{equation}
 {\mathbb A} = \bigcup_{{\mathcal P}} {\mathbb A} ({\mathcal P}),
 \ \ \ \  {\mathbb A} ({\mathcal P}) = {\mathbb R}\times \prod_{p\in
 {\mathcal P}} {\mathbb Q}_p
 \times \prod_{p\not\in {\mathcal P}} {\mathbb Z}_p \, .         \label{2.5}
\end{equation}

A multiplicative group of ideles $\mathbb{A}^\ast$ is a subset of
${\mathbb A}$ with elements $x= (x_\infty\,, x_2\,, x_3 \,, \cdots ,
x_p\,, \cdots)$ ,  where $x_\infty \in {\mathbb R}^\ast = {\mathbb
R} \setminus \{ 0\}$ and $x_p \in {\mathbb Q}^\ast_p = {\mathbb Q}_p
\setminus \{0 \}$ with the restriction that for all but a finite set
$\mathcal P$  one has that  $x_p \in {\mathbb U}_p$ . Thus the whole
set of ideles is
\begin{equation}
 {\mathbb A}^\ast = \bigcup_{{\mathcal P}} {\mathbb A}^\ast ({\mathcal P}),
 \ \ \ \ {\mathbb A}^\ast ({\mathcal P}) = {\mathbb R}^{\ast}\times \prod_{p\in {\mathcal P}}
 {\mathbb Q}^\ast_p
 \times \prod_{p\not\in {\mathcal P}} {\mathbb U}_p \, .         \label{2.6}
\end{equation}

A principal adele (idele) is a sequence $ (x, x, \cdots, x, \cdots)
\in {\mathbb A}$ , where $x \in  {\mathbb Q}\quad (x \in {\mathbb
Q}^\ast = {\mathbb Q}\setminus \{ 0\})$. ${\mathbb Q}$ and ${\mathbb
Q}^\ast$ are naturally embedded in  ${\mathbb A}$ and ${\mathbb
A}^\ast$ , respectively.

Let  ${\mathbb P}$ be set of all  primes $p$. Denote by
$\mathcal{P}_i \,, \,\, i \in \mathbb{N} ,$ subsets of ${\mathbb
P}$. Then ${{\mathcal P}_i} \prec {\mathcal P}_j$ if ${{\mathcal
P}_i} \subset {\mathcal P}_j$. It is evident that ${\mathbb A}({\cal
P}_i)\subset {\mathbb A}({\cal P}_j)$ when ${\mathcal P}_i \prec
{\mathcal P}_j$. Spaces ${\mathbb A}({\cal P})$ have natural
Tikhonov topology and adelic topology in ${\mathbb A}$ is introduced
by inductive limit: $ {\mathbb A} = \lim \mbox{ind}_{{\mathcal P}
\in {\mathbb P}} {\mathbb A}({\cal P})$. A basis of adelic topology
is a collection of open sets of the form $ W ({\mathcal P}) =
{\mathbb V}_\infty \times \prod_{p \in {\mathcal P}} {\mathbb V}_p\,
\times \prod_{p \not \in {\mathcal P}} {\mathbb Z}_p \, $, where
${\mathbb V}_\infty$ and ${\mathbb V}_p$ are open sets in ${\mathbb
R}$ and ${\mathbb Q}_p$ , respectively. Note that adelic topology is
finer than the corresponding Tikhonov topology. A sequence of adeles
$a^{(n)}\in {\mathbb A}$ converges to an adele $a \in {\mathbb A}$
if ${\it i})$ it converges to $a$ componentwise and ${\it ii})\, $
if there exist a positive integer $N$ and a set ${\mathcal P}$ such
that $\, a^{(n)}, \, a \in {\mathbb A} ({\mathcal P})$ when $n\geq
N$. In the analogous way, these assertions hold also for idelic
spaces ${\mathbb A}^\ast({\cal P})$ and ${\mathbb A}^\ast$.
${\mathbb A}$ and ${\mathbb A}^\ast$ are locally compact topological
spaces.

\section{\large Linear Fractional Dynamical Systems}

It is worth to recall some basic notions from the theory of
dynamical systems valid for mapping (\ref{1.1}) and its iterations
(\ref{1.2}) at real and $p$-adic spaces. Let us introduce an index
$v$ to denote  real ($v = \infty$) and $p$-adic ($v = p$) cases
simultaneously. A \textit{ fixed point} $\xi$ is a solution of the
equation $f (\xi) = \xi .$ If there exists a neighborhood $V (\xi)$
of the fixed point $\xi$ such that for any point $x_n \in V (\xi),
\,\, x_n \neq \xi$ holds: $(i) \,\, |x_n -\xi |_v < |x_{n-1} -
\xi|_v$, i.e. $\lim_{n \to \infty} x_n = \xi$, then $\xi$ is called
an \textit{attractor}; $(ii) \,\, |x_n -\xi |_v
> |x_{n-1} - \xi|_v$, then $\xi$ is  a \textit{repeller}; and $(iii)
\,\,  |x_n -\xi |_v = |x_{n-1} - \xi|_v $, then $\xi$ is  an
\textit{indifferent point}.
 Basin of attraction $A (\xi)$ of an attractor
$\xi$ is the set \beq A (\xi) = \{x_0 \in \mathbb{Q}_v : \lim_{n \to
\infty} x_n \to \xi \} .\label{3.1}\eeq
A Siegel disk is called a
ball $V_r (\xi)$ if every sphere $S_\rho (\xi), \, \rho < r$ is an
invariant sphere of the mapping $f (x)$, i.e. if an initial point
$x_0 \in S_\rho (\xi)$ then all iterations $x_n$ also belong to
$S_\rho (\xi)$. The union of all Siegel disks $V_r (\xi)$ with the
same center $\xi$ is called a maximum Siegel disk and denoted by $S
I (\xi)$.

  If the mapping (\ref{1.1}) has the first derivative in the fixed point $\xi$
then it is useful to employ the following properties: $|f' (\xi)|_v
> 1$ - attractor, $|f' (\xi)|_v < 1$ - repeller and $|f' (\xi)|_v = 1$
- indifferent point.

It is worth noting that the general form of a linear fractional
transformation is given by

\beq
 f (z) = \frac{a z + b}{c z + d}\,,  \label{3.2}
\eeq

\noi where $a, b, c, d, z \in \mathbb{C} \,\, \mbox{or} \,\,
\mathbb{C}_p$ with conditions $z \neq -\frac{d}{c}$, $c \neq 0$, $a
d - b c \neq 0 \,,$ and complex plane may be extended by the point
at infinity. There is an isomorphism between map (\ref{3.2}) and $2
\times 2$ matrices

\bea F =  \left(\begin{array}{ll}
 a  &  b  \\
 c &   d
                \end{array}
                \right)\,,
\quad \mbox{det} F \neq 0 \,,   \label{3.3}
                \eea
which are elements of $G L (2\,, \mathbb{C})$ or $G L (2\,,
\mathbb{C}_p)$ groups. Since map (\ref{3.2}) remains the same under
change $a \to \lambda a \,, b \to \lambda b \,, c \to \lambda c \,,
d \to \lambda d \,, $ one can choose a suitable $\lambda$ and
redefine $a\,, b\,, c\,, d$ so that $\mbox{det} F = a b - c d = 1 $.
Thus one obtains $S L (2\,, \mathbb{C})$ and $S L (2\,,
\mathbb{C}_p)$. If $a d - b c = 1$ the above map (\ref{3.2}) is
still invariant under transformation $a \to - a \,, b \to - b \,, c
\to - c \,, d \to - d \,, $ and the corresponding projective linear
groups are $ P S L (2\,, \mathbb{C})= S L (2\,, \mathbb{C}) / \{ \pm
E\}$ and $ P S L (2\,, \mathbb{C}_p)= S L (2\,, \mathbb{C}_p) /
\{\pm E\}$, where $E$ is  the unit $2 \times 2$ matrix.

We shall below mainly consider  rational dynamical systems given by
the following mapping

\beq
 f (x) = \frac{a x + b}{c x + d}\,, \quad x \in \mathbb{A} \,, \label{3.4}
\eeq

\noi where $a, b, c, d \in \mathbb{Q}$ with conditions $x \neq
-\frac{d}{c}$, $c \neq 0$ and $a d - b c = 1$. The corresponding
group of matrices $F\,,$  with $\, \mbox{det}\, F = 1$, is $S L (2,
\mathbb{Q})$, which is isomorphic to $Sp \,(2, \mathbb{Q})$ - the
group of symplectic $2 \times 2$ matrices with rational entries.

It is worth mentioning that the map (\ref{3.4}) preserves the
cross-ratio \beq   \frac{(\alpha_1 - \alpha_3) \, (\alpha_2 -
\alpha_4)}{(\alpha_1 - \alpha_4)\, (\alpha_2 - \alpha_3)}  =
  \frac{(f ( \alpha_1) - f (\alpha_3)) \, (f (\alpha_2 ) -
f (\alpha_4) )}{ (f (\alpha_1) - f (\alpha_4) )\, ( f(\alpha_2) - f
(\alpha_3 ))} \label{3.5} \eeq between any different points $x =
\alpha_1\,, \alpha_2\,, \alpha_3\,, \alpha_4\,$.

 To have an adelic system, it must be satisfied $|f_p (x_p)|_p \leq
1$ in
 \bea
f_{\mathbb{A}}(x) = \Big( f_\infty (x_\infty) \,, f_2 (x_2)\,, f_3
(x_3)\,, \cdots \,, f_p (x_p) \,, \cdots \Big) \,, \quad x \in
\mathbb{A}\,, \label{3.6} \eea

\noi for all but a finite set $\mathcal{P}$ of prime numbers $p$. In
other words, there has to be a prime number $q$ such that $|f_p
(x_p)|_p \leq 1$  for
 all $p > q$. It is obvious that $q$ depends not only on parameters
 $a, b, c, d \in \mathbb{Q}$, which characterize system, but also
 depends on $x \in \mathbb{Q}_p$.
 For this reason, let us consider existence of a prime number $q$
 such that

 \bea
   \Big| \frac{a x + b}{c x + d} \Big|_p \leq 1 \,,\quad c \neq 0 \,,
   \quad  d \neq 0\,, \quad x \in \mathbb{Z}_p \,, \label{3.7}
 \eea

\noi for all $p > q$. Requiring that $c$ and $d$ are nonzero
rational numbers then there exists an enough large prime number $q$
such that $ |c|_p = |d|_p = 1$, $|c x + d|_p = 1$ and $|a x + b|_p
\leq 1$ when $|x|_p \leq 1$ for all $p > q$. Then it follows the
existence of  prime $q$ such that (\ref{3.7}) is satisfied for all
but a finite set of primes, i.e. this function has necessary adelic
properties.  \vskip.4cm


For the function (\ref{3.2}) we find the following two fixed points:

\bea \xi_{1,2} = \frac{a -d \pm \sqrt{(a -d)^2 + 4bc}}{2c} \, =
\frac{a -d \pm \sqrt{(a + d)^2 - 4 ( ad - bc)}}{2c} \, \label{3.8}
\eea with properties \beq f(\xi_1)\cdot f(\xi_2) = \xi_1 \cdot \xi_2
= - \frac{b}{c} \, \,, \quad \quad f' (\xi_1) \cdot f' (\xi_2) = 1.
\label{3.9} \eeq  Let us note that (\ref{3.8}) can be rewritten in
the form \bea \xi_{1,2} =  \frac{a -d \pm \sqrt{(\mbox{Tr}\, F)^2 -
4\, \mbox{det} \, F}}{2c} \,, \label{3.10} \eea where $F$ is the
corresponding matrix (\ref{3.3}).

For the fixed points it is important to notice that if the point
$\xi_1$ is attractive ($|f' (\xi_1)|_v < 1$) then the point $\xi_2$
is repelling ($|f' (\xi_2)|_v > 1$) and vice versa. The indifferent
fixed points always emerge in the pair. These facts immediately
follow from the relation (\ref{3.9}) that holds for the mapping
associated with dynamical system, we consider.

 Generally, these
points belong to $\mathbb{C}$ in real case and $\mathbb{C}_p$ in
$p$-adic case, and their analysis we postpone for later
consideration. Now we are mainly interested in cases $a d - b c = 1$
and when fixed points are rational, because they then belong
simultaneously to real and $p$-adic numbers. To this end we are
going to analyze the following six possibilities: ({\bf A}) $b= 0$,
({\bf B}) $ b = c, \, d = a $, ({\bf C}) $b = -c\,, \, d = a + 2 c$,
({\bf D}) $ b = -c\,, \, d = a - 2 c $, ({\bf E}) $ d = - a + 2$ and
({\bf F}) $ d = - a - 2$ which give rational values of $\sqrt{(a
-d)^2 + 4bc} = \sqrt{(a +d)^2 - 4( a d - b c)}$ in (\ref{3.8}).

\vskip.3cm

\subsection{  Case A: $b =0,\,\, ad = 1 $.} We have the
following rational function with fixed points:

\bea f(x) = \frac{ x}{d (c x + d)} \,, \quad \xi_1 = \frac{1-d^2}{c
d} \,, \,\,\, \xi_2 =0\,, \,\,\, d \neq 0. \label{3.10} \eea

For further analysis we need

\bea f' (x) = \frac{1}{(c x + d)^2} \,, \quad f'(\xi_1) = d^2 \,,
\,\,\, f'(\xi_2) = \frac{1}{d^2} \,. \label{3.11} \eea We shall see
now that dynamics $ f(x) = \frac{ x}{d (c x + d)}$ has  attractive,
repelling and indifferent fixed points which do  not depend on
parameter $c\neq 0$, and that  there exists such finite set
$\mathcal{P}$ of primes   that $x_1$ and $x_2$ are indifferent
points for $p \not \in \mathcal{P}$.

\subsubsection{Fixed points in real and $p$-adic cases}

Here we have three distinct possibilities for fixed points $\xi_1
=\frac{1-d^2}{c d}$ and $\xi_2 =0$. Let $\mathcal{P}'= \big\{p \in
\mathbb{P} : |d |_p < 1 \big\}$  and $\mathcal{P}''=\big\{p \in
\mathbb{P} : |d |_p
        > 1 \big\}$, where $\mathbb{P}$ is the set of all prime numbers.

{\bf(i)}  Fixed point $\xi_1$ is \textit{attractive} ($|f'
(\xi_1)|_v <1$) and $\xi_2$ is \textit{repelling} ($|f' (\xi_2)|_v
>1$) iff $|d|_\infty < 1$ in the real case and $p \in \mathcal{P}'$
in the $p$-adic case.

\vskip.3cm

{\bf(ii)}  Fixed point $\xi_1$ is \textit{repelling} ($|f'
(\xi_1)|_v
>1$) and $\xi_2$ is \textit{attractive} ($|f' (\xi_2)|_v <1$) iff
$|d|_\infty > 1$ in the real case and $p \in \mathcal{P}''$ in the
$p$-adic case.

\vskip.3cm

{\bf(iii)}  Fixed points $\xi_1$  and $\xi_2$ are
\textit{indifferent} ($|f' (\xi_1)|_v = |f' (\xi_2)|_v =1$) iff
$|d|_\infty = 1$ in the real case and $p \not\in \mathcal{P}' \cup
\mathcal{P}''$ in the $p$-adic case.

\subsubsection{Adelic aspects of fixed points}

According to the above results one has the following two adelic
fixed points $\xi^{(i)}$ :

\bea \xi^{(i)} = \Big(\xi^{(i)}_\infty \,,
\xi^{(i)}_2\,,\xi^{(i)}_3\,,\xi^{(i)}_5\,, \cdots \,, \xi_p^{(i)}
\,, \cdots \Big) \,, \quad \xi^{(i)} \in \mathbb{A}\,, \label{3.12}
\quad i=1,2\,,\eea where $\xi^{(1)}_\infty = \xi^{(1)}_p =
\frac{1-d^2}{c d} \in \mathbb{Q}$ and $\xi^{(2)}_\infty =
\xi^{(2)}_p = 0$ for any $p \in \mathbb{P}$. Structure of adelic
fixed points $\xi^{(i)}$ depends only on the values of rational
parameter $d$. When $d =\pm 1$ both adeles $\, \xi^{(1)}$ and $\,
\xi^{(2)}$ have indifferent points in the real and all $p$-adic
positions, i.e. they are adelically indifferent. However, when $d
\neq \pm 1$ there emerge also attractive and repelling points in
real as well as in $p$-adic cases but only for finite sets of
primes. Namely if parameter $d \in \mathbb{Q}$ is such that
$|d|_\infty < 1$ (or $|d|_\infty
> 1$), $|d|_p < 1$ for $p \in \mathcal{P}'$ and $|d|_p > 1$ for
$p\in \mathcal{P}''$ then adelic fixed point $\xi^{(1)}$ has one
real attractive (or repelling) point and finite $p$-adic attractive
and repelling points which correspond to $\mathcal{P}'$ and
$\mathcal{P}''$, respectively. The vice versa situation is for the
second adelic point $\xi^{(2)}$.

The above results can be summarized as follows:

$$\xi^{(1)}_v= \frac{1-d^2}{cd}= \left\{\begin{array}{lll}
                 \mbox{attractive},  &  |d|_v<1  & ($p$\in \mathcal{P}')\\
                 \mbox{repelling},  &   |d|_v>1  & ($p$\in\mathcal{P}'')  \\
                 \mbox{indifferent},&   |d|_v=1  &
                 ($p$\not \in \mathcal{P}= \mathcal{P}'\cup\mathcal{P}'')\,,
                \end{array}
                \right.$$

$$\xi^{(2)}_v= 0=\left\{\begin{array}{lll}
                 \mbox{attractive},  &  |d|_v>1  & ($p$\in \mathcal{P}')\\
                 \mbox{repelling},  &   |d|_v<1  & ($p$\in\mathcal{P}'')  \\
                 \mbox{indifferent},&   |d|_v=1  &
                 ($p$\not\in \mathcal{P}=\mathcal{P}'\cup\mathcal{P}'')\,.
                \end{array}
                \right.$$

\bigskip

\subsection{ Case B:  $ c = b, \, d = a,\,\, a^2 - b^2 =1 $. }

In this case we have the following function with its fixed points:

\beq f(x) = \frac{ a x + b}{b x + a} \,, \quad \xi_1 = 1 \,, \,
\xi_2 = -1. \label{3.13}\eeq For the sequel we need \bea f' (x) =
\frac{1}{(a+b x )^2} \,, \quad f'(\xi_1) = \frac{1}{(a+b)^2} \,,
\,\,\, f'(\xi_2) = \frac{1}{(a-b)^2} \,. \label{3.14} \eea

\subsubsection{Fixed points in real and $p$-adic cases}

There are again three distinct possibilities. Let now $\mathcal{P}'=
\big\{p \in \mathbb{P} : |a-b |_p < 1 \big\}$  and
$\mathcal{P}''=\big\{p \in \mathbb{P} : |a-b |_p
        > 1 \big\}$.

{\bf(i)}  Fixed point $\xi_1$ is \textit{attractive} ($|f'
(\xi_1)|_v <1$) and $\xi_2$ is \textit{repelling} ($|f' (\xi_2)|_v
>1$) iff $|a-b|_\infty < 1$ in the real case and $p \in
\mathcal{P}'$ in the $p$-adic case.

\vskip.3cm

{\bf(ii)}  Fixed point $\xi_1$ is \textit{repelling} ($|f'
(\xi_1)|_v
>1$) and $\xi_2$ is \textit{attractive} ($|f' (\xi_2)|_v <1$) iff
$|a-b|_\infty > 1$ in the real case and $p \in \mathcal{P}''$ in the
$p$-adic case.

\vskip.3cm

{\bf(iii)}  Fixed points $\xi_1$  and $\xi_2$ are
\textit{indifferent} ($|f' (\xi_1)|_v = |f' (\xi_2)|_v =1$) iff
$|a-b|_\infty = 1$ in the real case and $p \not\in \mathcal{P}' \cup
\mathcal{P}''$ in the $p$-adic one.

\subsubsection{Adelic aspects of fixed points}

According to the above results one has the following two adelic
fixed points $\xi^{(i)}$ :

\bea \xi^{(i)} = \Big(x^{(i)}_\infty \,,
\xi^{(i)}_2\,,\xi^{(i)}_3\,,\xi^{(i)}_5\,, \cdots \,, \xi_p^{(i)}
\,, \cdots \Big) \,, \quad \xi^{(i)} \in \mathbb{A}\,, \label{3.15}
\quad i=1,2\,,\eea where $\xi^{(1)}_\infty = \xi^{(1)}_p = + 1$ and
$\xi^{(2)}_\infty = \xi^{(2)}_p = - 1$ for any $p \in \mathbb{P}$.
Structure of adelic fixed points $\xi^{(i)}$ depends only on the
rational values of  $a - b$. When $a - b = \pm 1$ both adeles $\,
\xi^{(1)}$ and $\, \xi^{(2)}$ have indifferent points in the real
and all $p$-adic positions, i.e. they are adelically indifferent.
However, when $a - b \neq \pm 1$ there emerge also attractive and
repelling points in real as well as in $p$-adic cases but only for
finite sets of primes. Namely, if  $a - b \in \mathbb{Q}$ is such
that $|a-b|_\infty < 1$ (or $|a-b|_\infty
> 1$), $|a-b|_p < 1$ for $p \in \mathcal{P}'$ and $|a-b|_p > 1$ for
$p\in \mathcal{P}''$ then adelic fixed point $\xi^{(1)}$ has one
real attractive (or repelling) point and finite $p$-adic attractive
and repelling points which correspond to $\mathcal{P}'$ and
$\mathcal{P}''$, respectively. The vice versa situation is for the
second adelic point $\xi^{(2)}$.

The above results can be summarized as follows:

$$\xi^{(1)}_v= 1 =\left\{\begin{array}{lll}
                 \mbox{attractive},  &  |(a-b)|_v<1  & ($p$\in \mathcal{P}')\\
                 \mbox{repelling},  &   |(a-b)|_v>1  & ($p$\in\mathcal{P}'')  \\
                 \mbox{indifferent},&   |(a-b)|_v=1  &
                 ($p$ \not \in\mathcal{P}'\cup\mathcal{P}'')\,,
                \end{array}
                \right.$$

$$\xi^{(2)}_v= -1 = \left\{\begin{array}{lll}
                 \mbox{attractive},  &  |(a-b)|_v>1  & ($p$\in \mathcal{P}')\\
                 \mbox{repelling},  &   |(a-b)|_v<1  & ($p$\in\mathcal{P}'')  \\
                 \mbox{indifferent},&   |(a-b)|_v=1  &
                 ($p$\not \in\mathcal{P}'\cup\mathcal{P}'')\,.
                \end{array}
                \right.$$

\vskip.3cm

\subsection{ Case C: $b = -c\,, \, d = a + 2 c \,, \,\, (a+c)^2
=1 $.}

This is the case with fused fixed points

\beq
 f (x) = \frac{a x - c}{c x + a + 2c}\,, \quad  \xi_1 = \xi_2 = -1 \,.
\label{3.16}\eeq For further investigation we need
 \bea f' (x) = \frac{1}{(c x +
a + 2c)^2} \,, \quad f'(\xi_1) =f'(\xi_2)= 1 \,. \label{3.17} \eea

\subsubsection{Fixed points in real and $p$-adic cases}

In this special case we have the only one possibility.  Namely, due
to $|f' (\xi_1)|_v = |f' (\xi_2)|_v =1$ it follows that the fused
fixed point  $\xi_1 = \xi_2 = -1 $ is indifferent one in real as
well as in all $p$-adic cases (i.e. for all primes $p \in
\mathbb{P}$).

\subsubsection{Adelic aspects of fixed points}

According to the above results one has only one adelic fixed point
$\xi^{(1)} = \xi^{(2)} \equiv \xi$, i.e.

\bea \xi = \big(\xi_\infty \,, \xi_2\,,\xi_3\,,\xi_5\,, \cdots \,,
\xi _p \,, \cdots \Big) \,, \quad \xi \in \mathbb{A}\,, \label{3.18}
\eea where $\xi_\infty = \xi_p = - 1$  for any $p \in \mathbb{P}$.
This is one pure adelic indifferent point for any rational values of
parameters $a$ and $c$ constrained by relation $(a +c)^2 = 1$ and $c
\neq 0$.

\vskip.3cm

\subsection{ Case D:  $b = -c\,, \, d = a - 2 c \,, \,\, (a-c)^2 = 1$.}

In this case one has again mapping with fused fixed points, i.e.

\beq
 f (x) = \frac{a x - c}{c x + a - 2c}\,, \quad  \xi_1 = \xi_2 = 1
 \,. \label{3.19}
\eeq In the following we need

\bea f' (x) = \frac{1}{(c x + a - 2c)^2} \,, \quad f'(\xi_1)
=f'(\xi_2)= 1\,. \label{3.20} \eea

\subsubsection{Fixed points in real and $p$-adic cases}

In this special case we have the only one possibility.  Namely, due
to $|f' (\xi_1)|_v = |f' (\xi_2)|_v =1$ it follows that the fused
fixed point  $\xi_1 = \xi_2 = 1 $ is indifferent one in real as well
as in all $p$-adic cases.

\subsubsection{Adelic aspects of fixed points}

According to the above results one has only one adelic fixed point
$\xi^{(1)} = \xi ^{(2)} \equiv \xi$, i.e.

\bea \xi = \big(\xi_\infty \,, \xi_2\,,\xi_3\,,\xi_5\,, \cdots \,,
\xi_p \,, \cdots \Big) \,, \quad \xi \in \mathbb{A}\,, \label{3.21}
\eea where $\xi_\infty = \xi_p =  1$  for any $p \in \mathbb{P}$.
This is one pure adelic indifferent point for any rational values of
parameters $a$ and $c$ constrained by relation $(a -c)^2 = 1$ and $c
\neq 0$.

\vskip.3cm

\subsection{ Case E: $d =-a+2\,, \,  \,\, (a -1)^2+ bc
=0 $.}

This is another case with double fixed point:

\beq
 f (x) = \frac{a x + b}{c x - a + 2}\,, \quad  \xi_1 = \xi_2 =\frac{a-1}{ c}
 \,. \label{3.22}
\eeq For further investigation we need
 \bea f' (x) = \frac{1}{(c x
-a + 2)^2} \,, \quad f'(\xi_1) =f'(\xi_2)= 1 \,. \label{3.23} \eea

\subsubsection{Fixed points in real and $p$-adic cases}

 Due to $|f' (\xi_1)|_v = |f' (\xi_2)|_v =1$ it follows that the
fused rational fixed point  $\xi_1 = \xi_2 = \frac{a-d}{2 c} $ is
indifferent one in real as well as in all $p$-adic cases.

\subsubsection{Adelic aspects of fixed points}

According to the above results we have only one adelic fixed point
$\xi^{(1)} = \xi^{(2)} \equiv \xi$, i.e.

\bea \xi = \big(\xi_\infty \,, \xi_2\,,\xi_3\,,\xi_5\,, \cdots \,,
\xi _p \,, \cdots \Big) \,, \quad \xi \in \mathbb{A}\,, \label{3.24}
\eea where $\xi_\infty = \xi_p = \frac{a-1}{ c}$  for any $p \in
\mathbb{P}$. This is one pure adelic indifferent point for any
rational values of parameters $a\,, b$ and $c$ constrained by
relation $ (a - 1)^2 + bc =0 $ and $c \neq 0$.

\vskip.3cm

\subsection{ Case F: $d =-a-2\,, \,  \,\, (a +1)^2 + bc
= 0 $.}

As in the previous three cases one has here fusion of fixed points.
Namely,

\beq
 f (x) = \frac{a x + b}{c x - a - 2}\, \quad  \xi_1 = \xi_2 =\frac{a + 1}{ c} \,.
\label{3.25}\eeq  We also employ
 \bea f' (x) = \frac{1}{(c x
-a - 2)^2} \,, \quad f'(\xi_1) =f'(\xi_2)= 1 \,. \label{3.26} \eea

\subsubsection{Fixed points in real and $p$-adic cases}

Since $|f' (\xi_1)|_v = |f' (\xi_2)|_v =1$ it follows that the fused
fixed point  $\xi_1 = \xi_2 = \frac{a-d}{2 c} $ is indifferent one
in real as well as in all $p$-adic cases.

\subsubsection{Adelic aspects of fixed points}

From the above results one has only one adelic fixed point
$\xi^{(1)} = \xi^{(2)} \equiv \xi$, i.e.

\bea \xi = \big(\xi_\infty \,, \xi_2\,,\xi_3\,,\xi_5\,, \cdots \,,
\xi _p \,, \cdots \Big) \,, \quad \xi \in \mathbb{A}\,, \label{3.27}
\eea where $\xi_\infty = \xi_p = \frac{a+1}{ c}$  for any $p \in
\mathbb{P}$. This is one pure adelic indifferent point for any
rational values of parameters $a$ and $c$ constrained by relation $
(a + 1 )^2 + bc = 0 $ and $c \neq 0$.

\section{\large Concluding Remarks}

Let us recall that in linear fractional function (\ref{3.4}) we have
considered parameters $a\,, b\,, c$ and $d$ as some rational
numbers. This is a natural requirement, since measured values of
physical quantities are rational, and it gives us also a possibility
to investigate the corresponding dynamical systems as the adelic
ones. To have in (\ref{3.4}) fixed points $\xi_1$ and $\xi_2$, which
depend on these parameters, rational we have made restriction on the
discriminant  $(a-d)^2 + 4bc$  so that it is a square and then
$\sqrt{(a-d)^2 + 4bc}$ is a rational, too. We have found six such
cases and investigated their real, $p$-adic and adelic properties in
detail. However in a more general setting $\sqrt{(a-d)^2 + 4bc}$ may
belong to the field $\mathbb{C}= \mathbb{R}(\sqrt{-1})$ of usual
complex numbers and the field $\mathbb{Q}_p(\sqrt{\tau})$ of
$p$-adic quadratic extensions or even $\mathbb{C}_p$ (see the Sec.
2). The most general case would be investigation of mapping
(\ref{3.2}). This is mathematically also interesting (cf. Ref.
\cite{mukhamedov1}) but its potential physical content should be
justified.

We have analyzed mapping (\ref{3.4}) with condition $a d - b c = 1$
which is related to  $S L (2, \mathbb{Q})$. Since function
(\ref{3.4}) and its derivative $f' (x) = (a d - b c)/ (c x + d)^2$
are invariant under scale transformation $ a \to \lambda a\,, b \to
\lambda b\,,  c \to \lambda c\,, d \to \lambda d  $ it follows that
any $a d - b c = r^2 \neq 0$ case can be transformed to $a d - b c =
1$ rescaling by $ \lambda = 1/r \in \mathbb{Q}$. As a consequence of
this scale invariance our evaluation can be easily extended to $a d
- b c = r^2  \in \mathbb{Q}^\ast$ case.

It is worth noting that our analysis in Sec. 3 contains also some
transformations related to the modular group, i.e. transformations
of the form (\ref{3.4}) with $a\,, b\,, c\,, d \in \mathbb{Z}$
satisfying $a d - b c = 1$. Modular group identifies with quotient
group $S L (2, \mathbb{Z})/\{ \pm E\} $, where $E$ is the unit $2
\times 2$ matrix. We found the following five different such
transformations: \bea f_1 (x) = \frac{\pm x}{ c x \pm 1}\,, \quad
f_2 (x) = \frac{a x + a \mp 1}{(- a \pm 1) x - a \pm
2} \,, \quad f_3 (x) = \frac{(- c \pm 1 ) x - c}{ c x + c \pm 1} \,, \label{4.1} \\
f_4 (x) = \frac{( a x - a \pm 1)}{ ( a \mp 1) x - a \pm 2} \,, \quad
f_5 (x) = \frac{(c \pm 1) x - c}{c x - c \pm 1} \, \,. \qquad
 \label{4.2}\eea

One can introduce product of norms on the multiplicative group of
ideles $\mathbb{A}^\ast$ as \beq |\alpha|=\prod_{v}|\alpha_v|_{v}
\,,\quad \alpha = \{\alpha_v\}_v \in \mathbb{A}^\ast \,, \label{4.3}
\eeq where $v$ runs through primes $p$ and $\infty$. Here the
product in the right hand side makes sense because $|\alpha_p|_p =
1$ for  all but a finite set of $p\in\mathbb{P}$. This yields to the
remarkable adelic product formula  \beq |r| = |r|_\infty \prod_{p\in
\mathbb{P}}|r|_{p}= 1, \quad r\in\mathbb{Q}^\ast \,. \label{4.4}
\eeq This product formula connects real and all $p$-adic norms of
the same nonzero rational number. It presents the simplest example
of adelic product formulas which connect  real and $p$-adic
counterparts of a rational quantity. Note that $|r|^{s + i t} =1$,
where $s\,, t \in \mathbb{R}$, is a slight extension of (\ref{4.4})
and is an example of adelic  multiplicative  character. It is
obvious that  $r$ in (\ref{4.4})  can be replaced by nonzero
parameters $a\,, b\,, c\,, d$ as well as by fixed points $\xi_1\,,
\xi_2$. It can be also applied to the invariant relation (\ref{3.5})
and to the function $f (x)$ defined in (\ref{3.4}), when $x \in
\mathbb{Q}^\ast$. One of the important consequences of adelic
product formula (\ref{4.4}) is that a real quantity can be expressed
as product of inverse $p$-adic counterparts.

\bigskip

\section*{\large Acknowledgments}

The work on this article was partially supported by the Ministry of
Science and Environmental Protection, Serbia, under contract No
144032D. This research was also supported by the grant of  Swedish
Royal Academy of Science and the grant on p-adic dynamical systems
of the International Center for Mathematical Modeling in Physics,
Engineering, Economy and Cognitive Science, University of V\"axj\"o,
Sweden.

\end{document}